\documentclass[aip,reprint]{revtex4-1}

\usepackage{amsmath,amsthm,amssymb}
\usepackage{graphicx}

\usepackage{dcolumn}
\usepackage{bm}
\usepackage{color}
\usepackage{epsfig}
\usepackage{multirow}
\usepackage{mathrsfs}
\usepackage{hyperref}
\usepackage{cleveref}
\usepackage{epstopdf}
\usepackage{subfigure}
\usepackage{autobreak}

\newcommand{\V}[1]{\boldsymbol{#1}} 
\newcommand{\M}[1]{\boldsymbol{#1}} 
\newcommand{\D}[1]{\Delta#1} 
\renewcommand{\d}[1]{\delta#1} 
\newcommand{\abs}[1]{\left|#1\right|} 

\newcommand{\grad}{\M{\nabla}} 

\newcommand{\eps}{\epsilon}
\usepackage{color}
\newcommand{\rev}[1] {{\color{black}#1}}

\usepackage[utf8]{inputenc}
\usepackage[T1]{fontenc}
\usepackage{mathptmx}
\usepackage{etoolbox}

\makeatother
\begin{document}


\title{Broken Symmetries in Quasi-2D Charged Systems via \rev{Negative} Dielectric Confinement} 



\author{Xuanzhao Gao}
\email[]{xz.gao@connect.ust.hk}
\affiliation{Thrust of Advanced Materials, The Hong Kong University of Science and Technology (Guangzhou), Guangdong, China}
\affiliation{Department of Physics, The Hong Kong University of Science and Technology, Hong Kong SAR, China}

\author{Zecheng Gan} \thanks{Corresponding author}
\email[]{zechenggan@ust.hk}
\affiliation{Thrust of Advanced Materials, The Hong Kong University of Science and Technology (Guangzhou), Guangdong, China}
\affiliation{Department of Mathematics, The Hong Kong University of Science and Technology, Hong Kong SAR, China}


\date{\today}

\begin{abstract}
    \rev{We report spontaneous symmetry breaking (SSB) phenomena in symmetrically charged binary particle systems under planar nanoconfinement with negative dielectric constants}. 
    The SSB is triggered $solely$ via the \rev{dielectric confinement effect}, without any external fields.
    The mechanism of SSB is found to be caused by the strong polarization field enhanced by nanoconfinement, giving rise to charge/field oscillations in the transverse directions.
    Interestingly, dielectric contrast can even determine the degree of SSB in transverse and longitudinal dimensions, forming charge-separated interfacial liquids and clusters on square lattices.
    Furthermore, we analytically show that the formed lattice constant is determined by the dielectric mismatch and the length scale of confinement, which is validated via molecular dynamics simulations.
    The novel broken symmetry mechanism may provide new insights in the study of quasi-2D systems and the design of future nanodevices.
\end{abstract}


\maketitle 
\textit{Introduction.}--Quasi-2D systems are attracting much attention because of their huge potential in future nanodevices. Typically, such systems possess a nano-sized longitudinal thickness in the~$z$ direction, achieved through confinement, 
bulk-like and modeled as periodic in the transverse~$xy$ directions~\cite{mazars2011long}. 
Rich new collective behaviors arise in such systems, to name a few, polyelectrolyte adsorption and structure~\cite{messina2004effect,yuan2020structure}, ion transport and selectivity~\cite{nishizawa1995metal,cervera2006ionic}. 

Nevertheless, in terms of the spontaneous symmetry breaking (SSB) phenomena, much existing study focuses on purely 2D and 3D systems~\cite{levin:PRL:2008,joyce2011quasistationary,pakter2018nonequilibrium}, far less is known about quasi-2D.
For bulk electrolytes or neutral plasma, it is well-known that the Coulomb potential can be dynamically screened by surrounding countercharges, leading to effectively short-range interacting particle systems~\cite{huckel1923theory}. 
The situation becomes very different in quasi-2D charged systems: their reduced symmetry (i.e., the nano-sized confinement) weakens the electrostatic screening, and correlation effect can become much more important. Clearly, this is quasi-2D specific, where simplified 2D description would fail. 
Yet, to the best of our knowledge, no SSB phenomena have been reported in suspension of charge- and size-symmetric, overall-neutral particle systems under dielectric confinement, without any external fields.

Another important effect associated with quasi-2D charged systems concerns the permittivity, i.e., the \emph{dielectric confinement effect}. 
Substrate materials used for nanoscale confinement can range from dielectric to metallic, and nowadays, electromagnetic metamaterials, which have been developed with permittivities that can take negative values~\cite{veselago1967electrodynamics, smith2004metamaterials} when excited by electromagnetic waves of specific frequencies. 
Great efforts have been made to develop negative permittivity materials in the low frequency limit~\cite{cheng2017tunable, xie2022recent, xu2020polyaniline}. 
\rev{Noteworthy is that materials with \emph{negative static permittivity}  has drawn a considerable attention, though rarely seen, its existence has been predicted in materials such as metals and non-ideal plasma~\cite{Dolgov1918admissible,homes2001optical}. More recently, it has been experimentally achieved in a wide range of materials such as VO$_2$ films~\cite{kana2016thermally}, graphene~\cite{nazarov2015negative}, nanocolloids~\cite{shulman2007plasmalike}, and polymeric systems~\cite{yan2013negative}. }
\rev{Interestingly, even for water, the perpendicular component of its tensorial dielectric function has been observed to be negative within sub-Angstrom distances from the surface by nano confinement~\cite{Kornyshev1996Static, Schlaich2016Water, Kornyshev2021Nonlocal}.}

The confinement effect turns out to be physically interesting even when only a \emph{single} dielectric substrate is present. For electrolytes/polymers near a single
dielectric substrate, recently calculations have revealed
that the dielectric surface effect can significantly deviate the systems from bulk behaviors, examples include ion transport~\cite{antila2018dielectric}, polymer brush structure~\cite{yuan2020structure}, and pattern formation in dipolar films~\cite{wang2019dielectric}, where such effect is particularly enhanced when the substrate's permittivity is negative.
Unfortunately, incorporating a second dielectric substrate in the models to actually achieve dielectric confinement in computer simulations is far from straightforward. 
\rev{Although simulation techniques~\cite{arnold2002electrostatics, de2002electrostatics,tyagi2007icmmm2d,fernandez2010collection,jadhao2012simulation,zwanikken2013tunable,fahrenberger2014computing,dos2017simulations,yu2018plasmonic,liang2020harmonic,yuan2021particle,maxian2021fast} have made significant progress over the past decades, accurate and efficient treatment of the dielectric confinement effect remains challenging, especially when the system is strongly confined or substrates are with negative permittivity.}

In this work, through computer simulations of a prototypical charge- and size-symmetric binary mixture of particles
described by the primitive model~\cite{mcmillan1945statistical}, we demonstrate
that broken symmetries arise spontaneously due to the dielectric confinement effect alone. 
Moreover, we discover that the substrates permittivity can even qualitatively
alter the degree of SSB in transverse and  longitudinal
dimensions, forming charge-separated interfacial liquids and clusters on square lattices.
The mechanism of SSB is found to be caused by the strong polarization field enhanced by dielectric confinement, giving rise to charge/field oscillations in
the transverse directions.
It is discovered that the formed lattice constant can be quantitatively determined
by the dielectric mismatch and the length scale of confinement, which is analyzed theoretically, and also validated numerically via molecular dynamics simulations under various system settings.

\begin{figure}[htbp]
	\centering
	\includegraphics[width=0.45\textwidth]{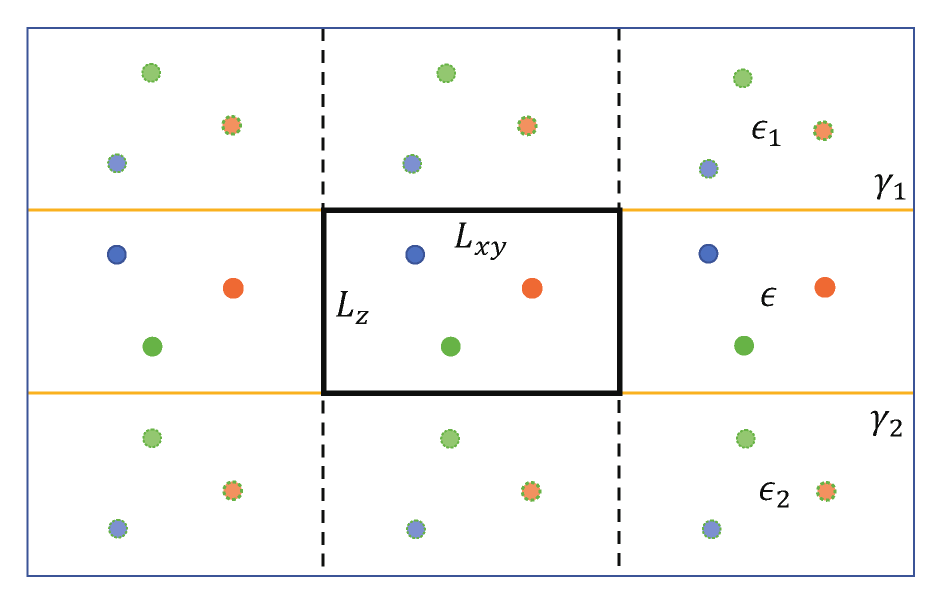}
	\caption{
		The figure illustrates a quasi-2D charged system and depicts the dielectric confinement effect from the viewpoint of the Image Charge Method (ICM). 
        The solvent medium, with dielectric permittivity~$\epsilon$, is represented by the middle layer, while the upper and lower layers represent the substrate with dielectric permittivities of~$\epsilon_1$ and~$\epsilon_2$, respectively. 
        The real charged particles of the doubly-periodic system are represented by colored circles surrounded by solid lines. 
        The dotted lines represent the image charges that are reflected by the dielectric interfaces in the~$z$ direction.
		\label{fig:ICM}
	}
\end{figure}

\textit{Model.}--The modeled geometry of the dielectric confined quasi-2D systems used for simulations is presented in Fig.~\ref{fig:ICM}. 
The system is doubly-periodic in the transverse direction and finite in the longitudinal direction, with edge lengths of~$L_x$,~$L_y$, and~$L_z$. 
All charged particles are located between the dielectric substrates with dielectric permittivity~$\eps_1$ and~$\eps_2$ and are immersed in a solvent with dielectric permittivity~$\eps$. 
Based on the ICM, the strength of polarization is quantified by the dimensionless coefficient reflection rates,~$\gamma_1$ and~$\gamma_2$, which are given by~$(\eps - \eps_i)/(\eps + \eps_i)$. 
The Green's function of Poisson's equation in such systems can be constructed via a multiple reflection process, resulting in an infinite image charge series as schematically illustrated in Fig.~\ref{fig:ICM}. 
Note that when~$\abs{\gamma_1 \gamma_2}\leq 1$, the image reflection series is convergent, but when~$\abs{\gamma_1 \gamma_2}>1$, it becomes~\emph{divergent}, and the reflective ICM approach fails. 
Therefore, current simulation studies in the~$\abs{\gamma}\geq 1$ regime are limited to a single dielectric substrate~\cite{wang2019dielectric}. 
However, our new approach overcomes this divergence issue using a proper renormalization strategy, allowing us to explore the dielectric confinement effect in all possible~$\gamma$ regimes, particularly the less explored scenario of metamaterial substrates with static negative permittivity.

The Green's function~$G(\V{r},\V{s})$ for Poisson's equation in a dielectric confined quasi-2D system can be expressed as 
\begin{equation}
    -\grad\cdot\left[\eta(\V r) \grad G(\V{r},\V{s})\right] = 4 \pi \d(\V r - \V s )\;,\label{eq:Green4Poisson}
\end{equation}
where~$\V{r}$ and~$\V{s}$ denote the target and source locations within the confined geometry,
and the relative dielectric function~$\eta(\V r) = \eps(\V{r}) / \eps$, where $\eps(\V{r})$ is a material-specific, piece-wise constant, defined as
\begin{equation}
    \eps(\V{r}) = \left\{
    \begin{array}{cc}
        \eps_1,~ & z > L_z \\
        \eps,~ & 0 \leq z \leq L_z \\
        \eps_2,~ & z < 0
    \end{array}
    \right. \;,
\end{equation}
and depicted in Fig.~\ref{fig:ICM}. 
Though such homogeneous dielectric constant approximation is a commonly used coarse-grained strategy in classical molecular dynamics (MD), it should be noted that for real materials, the dielectric function $\eps(\V{r})$ can be spatially varying with charge concentrations~\cite{Hasted2004Dielectric,Lyashchenko1998Complex}, wave lengths~\cite{Dolgov1918admissible,Kornyshev1996Static}, or local electric fields~\cite{Booth1951Dielectric,Booth1955Dielectric}. 
\rev{Moreover, for systems under aqueous nanoconfinements, the dielectric constant of water can become anisotropic and should be modelled as tensorial near the confinement surfaces~\cite{Loche2020universal}.}
Such dielectric variation effect has been less studied if coupled with particle-based simulations, but is important in understanding physical properties at finer time/length scales.

Finally, the dielectric interface conditions require that~$G(\V r, \V s)$ and~$\epsilon(\V r)\partial_z G(\V r, \V s)$ be continuous across~$z=0$ and~$L_z$, with the free-space boundary condition (FBC) holding as~$z\to\pm\infty$. 
It should be noted that proposing the proper FBC for charges under dielectric confinement requires careful consideration to ensure it is physically well-defined, and this will be clarified later.
In our discussion, we fix~$\eps=1$ for simplicity, and~$\eps_1=\eps_2=\eps'$, so that~$\gamma_1=\gamma_2=\gamma$. 
By varying~$\eps'$, we can change~$\gamma$ from~$-10$ to~$10$. 
\rev{As a possible experimental realization, the permittivity of VO$_2$ film in the long wavelength limit is approximately~$-14$ at~$350$K~\cite{kana2016thermally}. 
By choosing appropriate solvents with permittivities of approximately~$11.4$ or~$17.1$ (such as organic solvents), the proposed~$\gamma$ regimes can be achieved.}
Note that for general dielectric confinement setups with realistic dielectric constants, it is always possible to rescale the dielectric constants by $\eps$ and the confined charge densities by $\sqrt{\eps}$, so that the electrostatic system is mathematically equivalent. 

\rev{
\textit{Numerical method.}--To solve the long-range Coulomb interaction of quasi-2D charged systems with dielectric confinement, we proposed the following method.

First, following the work of A. P. Dos Santos et al.~\cite{dos2017simulations}, plane wave expansion is applied on both sides of Eq.~\eqref{eq:Green4Poisson}, which gives
\begin{equation}\label{eq:G_point_charge}
    \begin{split}
        G(\V{r},\V{s}) & = - \frac{1}{\pi} \iint_{\mathbb{R}^2} g(k, z, z_s) e^{-i \V{k} \cdot \Delta \V{\rho}} \text{d} k_x \text{d} k_y \\
        & = - \int_{0}^{+\infty} 2 g(k, z, z_s) J_0(k \Delta \rho) k \text{d}k\;,
    \end{split}
\end{equation}
where~$\V{k} = (k_x, k_y)$,~$\Delta \V{\rho} = (x - x_s, y - y_s)$. For~$k > 0$, by applying the Dirichlet-to-Neumann map, it can be shown that~$g(k, z, z_s)$ satisfies the following 1D boundary value problem,
\begin{equation}
    \begin{split}\label{eq:g_ODE}
        \frac{\partial^2 g(k, z, z_s)}{\partial z ^2} - k^2 g(k, z, z_s) &= \delta(z - z_s)\;,\\
        \eps \partial_z g(k, L_z, z_s) + \eps_1 k g(k, L_z, z_s) &= 0 \;,\\
        \eps \partial_z g(k, 0, z_s) - \eps_2 k g(k, 0, z_s) &= 0\;.
    \end{split}
\end{equation}
The solution is given as
\begin{equation}\label{eq:g_solution}
    g(k, z, z_s) = \frac{1}{2k} \frac{1}{\gamma_1 \gamma_2 \exp{(-2 k L_z)} - 1} \sum_{i = 1}^{4} \Gamma_l \text{e}^{-k a_l}\;,
\end{equation}
where~$\Gamma_l = \left[1, ~\gamma_1, ~\gamma_2, ~\gamma_1 \gamma_2 \right]$ and~$a_l = [\abs{z - z_s}, ~z + z_s, ~2L_z - (z + z_s), ~2L_z - \abs{z - z_s}] \in [0, 2L_z]$.
And for~$k = 0$, the solution is given by
\begin{equation}\label{eq:g_k=0}
    g(k = 0, z, z_s) = - \frac{\abs{z - z_s}}{2}.
\end{equation}
Physically, Eq.~\eqref{eq:g_k=0} implies that for $k=0$, confined source charge acts as a uniformly charged plate.

Then to efficiently handle the electrostatic interaction, we develop a novel, modified Ewald splitting technique, which reads
\begin{equation}\label{eq:split}
    \delta(\V r) = \left[\d(\V r) - \frac{\alpha}{\pi} e^{-\alpha \rho^2}\d(z)\right] + \frac{\alpha}{\pi} e^{-\alpha \rho^2}\d(z)\;,
\end{equation}
with~$\V{\rho} = (x, y)$,~$\rho = \sqrt{x^2+y^2}$, and the choice of $\alpha$ will be determined by considerations of computational efficiency.
Same as the traditional Ewald splitting, subtracting and adding the Gaussian cloud splits the electrostatic interaction into short- and long-range components, which now converges rapidly in real and reciprocal spaces, respectively. 
Notice that the splitting strategy we propose here is tailored for the quasi-2D geometry, i.e.,
it avoids the subtle situation of Gaussian charge cloud overlapping the substrates.
Due to the splitting strategy Eq.~\eqref{eq:split}, the doubly-periodic Green's function can be decomposed into short- and long-range components, i.e., $G_1$ and $G_2$, satisfying
\begin{equation} \label{eq:G12}
    \begin{split}
        -\grad^2 G_1(\V r, \V s) &= 4 \pi \left[\d(\V r - \V s) - \frac{\alpha}{\pi} \d(z - z_s) e^{-\alpha \Delta \rho^2} \right]\;,\\
    	-\grad^2 G_2(\V r, \V s) &= 4 \pi \frac{\alpha}{\pi} \sum_{\V{m}} \d(z - z_s) e^{-\alpha \Delta \rho_{\V{m}}^2}\;,
    \end{split}
\end{equation}
where~$\V{m} = (m_x, m_y) \in \mathbb{Z}^2$ is the index of doubly-periodic images, and~$\Delta \rho_{\V{m}} = (x - x_s + L_x m_x, y - y_s + L_y m_y)$.
Eq.~\eqref{eq:G12} can be solved by convolution of the charge density over~$g(k, z, z_s)$, which gives
\begin{equation}\label{eq:G12_int}
    \begin{split}
        G_1(\V r, \V s)
        = & - \int_{0}^{+\infty} 2 g(k, z, z_s) (1-e^{-\frac{\V{k}^2}{4 \alpha}}) J_0(k \Delta \rho) k \text{d} k \;, \\
        G_2(\V r, \V s)
        = & - \frac{4 \pi}{L_x L_y}  \sum_{\V{k}} g(k, z, z_s) e^{-\frac{\V{k}^2}{4 \alpha}}e^{\mathrm{i} \V{k} \cdot \Delta \V \rho_m}\;.
    \end{split}
\end{equation}
Due to our splitting strategy,~$G_1$ and~$G_2$ decay rapidly in real and reciprocal space, respectively.
Thus one can simply apply numerical cutoff in real and reciprocal space with parameters~$r_c$ and~$k_c$. (For our numerical scheme and its error estimates, and also details about the renormalization technique when $|\gamma|>1$, and efficient implementation in computing $G_1$~\cite{trefethen2022exactness}, one can refer to the supplementary material~\cite{SI}.)
Finally, the total electrostatic energy can be calculated as
\begin{equation}
    \begin{split}
        U_{ele}
        = & - \sum_{\Delta \rho_{ij} < r_c} q_i q_j \int_{0}^{+\infty} g(k, z_i, z_j) (1-e^{-\frac{\V{k}^2}{4 \alpha}}) J_0(k \Delta \rho_{ij}) k \text{d} k \\
        & - \frac{2 \pi}{L_x L_y} \sum_{k<k_c} \sum_{i,j = 1}^N q_i q_j g(k, z_i, z_j) e^{-\frac{\V{k}^2}{4 \alpha}}e^{\mathrm{i} \V{k} \cdot \Delta \V{\rho}_{ij}}\;.
    \end{split}
\end{equation}
}

\begin{figure}[htbp]
	\centering
	\includegraphics[width=0.45\textwidth]{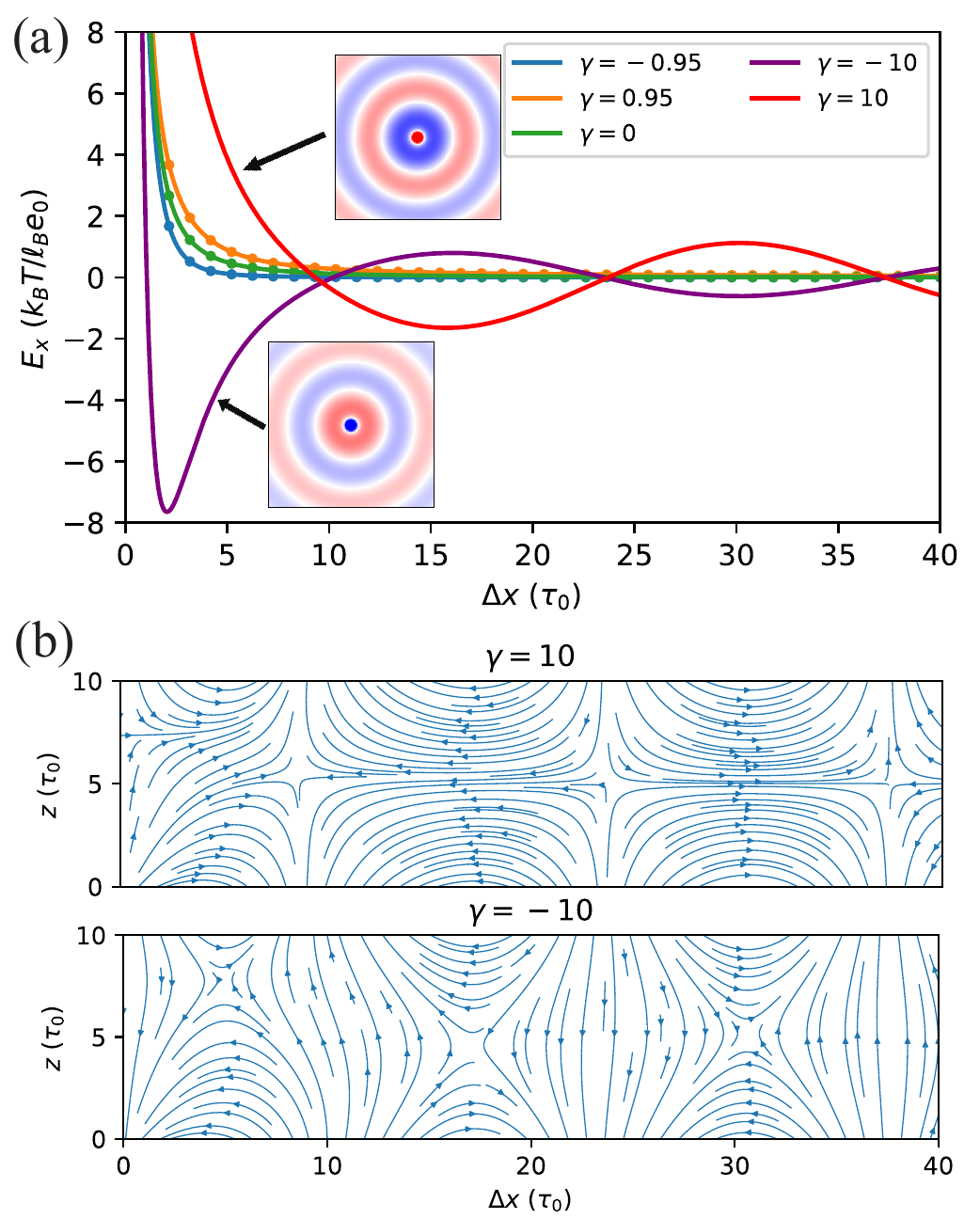}
	\caption{(a): the electric fields along~$x$ direction, generated by a cation with valence~$\nu=1$, fixed at~$z=\tau_0$, and confined by a pair of dielectric substrates located at~$z=0$ and~$10\tau_0$. Subplots depict the polarization charge density on the lower substrates. 
    (b): the corresponding field lines for the~$\gamma=\pm10$ scenarios.
		\label{fig:force_x}
            }
\end{figure}

\textit{Oscillatory single particle field.}--The dielectric confinement effect turns out to be physically fascinating even in the presence of a single charged particle. 
In Fig.~\ref{fig:force_x} (a), we present the electric field in the $x$ direction generated by a cation with valence~$\nu=1$ located at~$(x_0, y_0, \tau_0)$ in a quasi-2D system with a thickness of~$10\tau_0$, as a function of the distance from the cation~$\Delta x=x-x_0$, for different reflection rates~$\gamma$ characterizing the confinement. 
The field is defined as~$-\nu\ell_B\partial_x G(\mathbf{r}, \mathbf{r_0})$, where~$G(\mathbf{r}, \mathbf{r_0})$ is given by Eq.~\eqref{eq:G_pv}, and~$\ell_B=e_0^2/(4\pi\epsilon_0\epsilon k_B T)$ is the Bjerrum length of the solvent, with~$e_0$ the elementary charge,~$\epsilon_0$ the vacuum permittivity,~$k_B$ the Boltzmann constant, and~$T$ the temperature. 
For~$\vert\gamma\vert<1$ cases, as illustrated by the blue ($\gamma=-0.95$) and orange ($\gamma=0.95$) lines in Fig.~\ref{fig:force_x} (a), the polarization weakens or enhances the bare Coulomb field ($\gamma=0$), but with no qualitative difference. 
The results obtained by our method are in good agreement with those obtained by ICM, shown in dots in Fig.~\ref{fig:force_x} (a). 
However, for~$\vert\gamma\vert>1$, the results become non-trivial and qualitatively different. 
At short distance ($\tau_0<\Delta x < 10\tau_0$), we observe from Fig.~\ref{fig:force_x}(a) a continuous transition in the the near field interaction from like-charge attraction (LCA) into repulsion as~$\gamma$ increases from $-10$ to $+10$,
which can be understood as a significant enhancement of the polarization effect for~$\abs{\gamma}<1$ cases.
Even more interesting is the far field, it no longer decays monotonically but exhibits oscillatory behavior, which is rarely reported in previous studies.

\begin{figure}[htbp]
    \centering
    \includegraphics[width=0.45\textwidth]{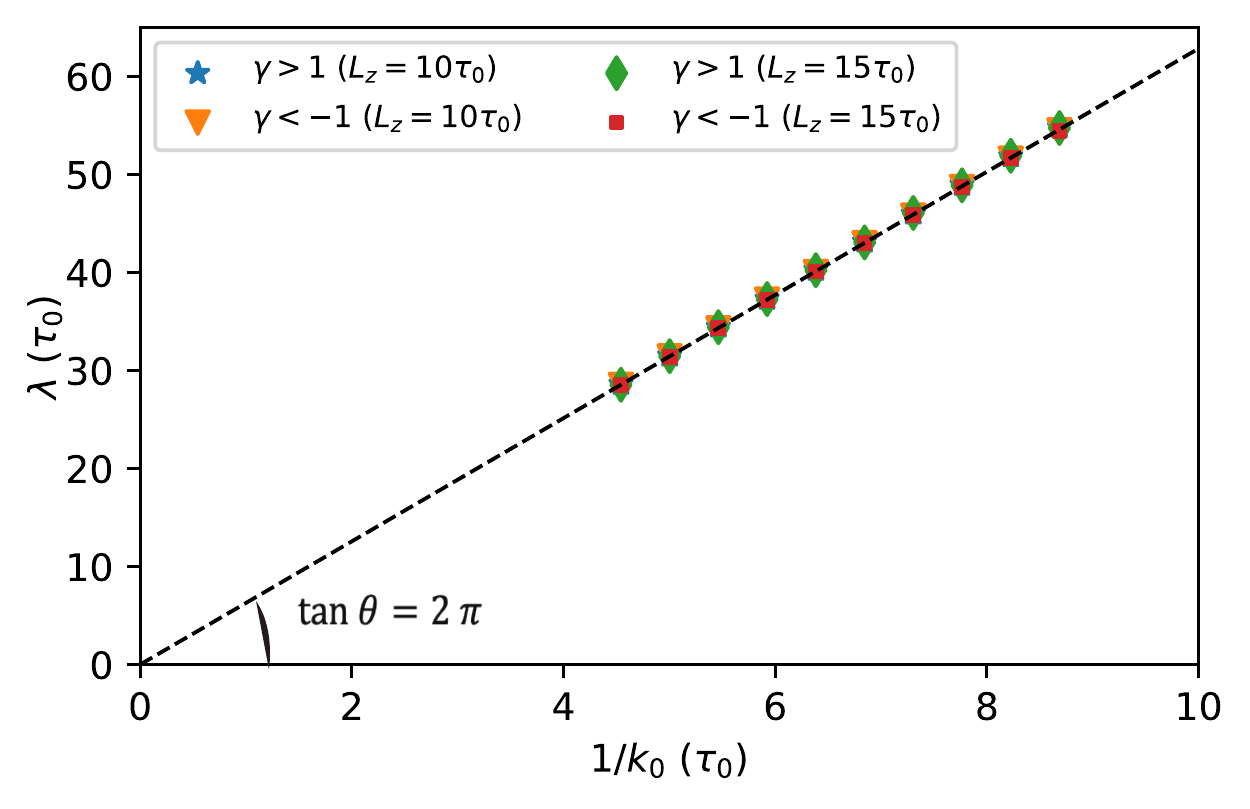}
    \caption{
        Numerical validations for the relationship between~$k_0$ and~$\lambda$ under various system parameter settings of~$\gamma$ and~$L_z$. For each case, $\lambda$ is approximated by averaging distances between nearby zeros of~$E_x$, and with different (randomly generated) locations in~$z$. 
    }
    \label{fig:k_wavelegth}
\end{figure}

To understand the origin of field oscillations, the polarization charge density profile on the substrate at~$z=0$ is shown in the subplots of Fig.~\ref{fig:force_x} (a). 
The charge density is defined by 
\begin{equation}
    \sigma(\V{r}) = \lim_{z \to 0^+} \nu \ell_B \eps_0  \left( 1 - \frac{\eps}{\eps'} \right) \partial_z G(\V{r}, \V{r_0})\;,
\end{equation}
and the field lines generated by $\sigma(\V r)$ are sketched in Fig.~\ref{fig:force_x} (b). 
The field oscillation is found to be generated by the strong transverse polarization charge density waves, influencing both the near and far fields. 
The oscillatory field lines has a very similar structure to that of a surface plasmonic resonance wave~\cite{willets2007localized}, but the physical origin is different. 
The oscillation is due to the reflected polarization enhanced by the dielectric confinement, characterized by parameters~$\gamma_1$,~$\gamma_2$, and~$L_z$. Particularly,
The confinement induced oscillation wave number is given by
\begin{equation}\label{eq:k0}
    k_0 = \frac{\ln{\gamma_1 \gamma_2}}{2 L_z}\;,
\end{equation}
which we will show analytically that this corresponds to a first-order pole in the Sommerfeld integral representation of the Green's function.
And the wavelength of the oscillation, defined as two times the distance between nearby zeros, satisfies 
\begin{equation}\label{eq:wavelength}
    \lambda \cdot k_0 = 2 \pi \;.
\end{equation}
Numerical validation shows that Eq.~\eqref{eq:wavelength} is highly robust under different choices of~$\V{r}$,~$\V{r}_0$,~$\gamma$, or~$L_z$, as shown in Fig.~\ref{fig:k_wavelegth}.
Importantly, the oscillation fields can be accurately predicted and controlled by adjusting~$k_0$. 
Eq.~\eqref{eq:k0} also indicates that the oscillation shall be weakened as $L_z$ is increased, and becomes non-oscillatory when $\gamma_1\gamma_2<1$.

\textit{Theoretical origin for oscillations.}--
\rev{Eq.~\eqref{eq:G_point_charge} shows that the Green's function can be represented as a Sommerfeld integral, and the analytical form of $g(k, z, z_s)$ indicates that it has non-trival behaviors.}
Clearly, $g(k, z, z_s)$ is divergent at $k=k_0$ (given in Eq.~\eqref{eq:k0}), and as $\gamma_1\gamma_2$ increases to be larger than 1, $k_0$ will shift onto the positive real axis, then the Sommerfeld integral needs to be renormalized.
Notice that when~$k \to k_0$, the divergent factor has the property
\begin{equation}
    \frac{1}{\gamma_1 \gamma_2 \exp{(-2 k L_z)} - 1} \to \frac{1}{2 L_z (k_0 - k)}\;,
\end{equation}
so that~$k_0$ is a first-order pole and the Cauchy principal value exists.
Then Eq.~\eqref{eq:G_point_charge} for~$\gamma_1 \gamma_2 > 1$ cases is given by
\begin{equation}\label{eq:G_pv}
    G(\V{r}, \V{s}) = - \text{p.v.} \left[ \int_{0}^{+\infty} 2 g(k, z, z_s) J_0(k \Delta \rho) k \text{d}k \right]  \;,
\end{equation}
which can be calculated numerically. In what follows, we analyze the oscillatory behavior~(for more details, see Supplementary Information (SI)~\cite{SI}). First, the Green's function consists of integrals of the following general form
\begin{equation}
    I_o = \int_0^{\infty} \frac{J_0(k \D \rho) \text{e}^{-ka}}{\exp{\left( 2 L_z (k_0 - k) \right)} - 1} \text{d}k\;,
\end{equation}
where~$\D \rho$,~$k_0$ and~$a$ are all positive constants.
We find that~$I_o$ can be further expanded as
\begin{equation}
    I_{o} = \frac{e^{-k_0 a}}{2L_z} \int_0^{\infty} \frac{J_0(k^\prime)}{k_0 \D \rho - k^\prime} \text{d}k^\prime + f(k_0, \D \rho, a),
\end{equation}
where~$k^\prime = k \D \rho$, and~$f(k_0, \D \rho, a)$ is a non-oscillatory analytic function which has minor contribution to~$I_o$.
The first integral term can be understood as a function of~$k_0 \D \rho$, or denoted as~$I_{m} (k_0 \D \rho)$. Clearly, $I_{m}$ is solely controlled by~$k_0$, given different parameters of $\gamma$ and $L_z$.
It is found that the first-order pole in $I_{m}$ provides the oscillatory mode, and we also numerical validated that the wavelength of the oscillation in $I_{m}$ indeed satisfies Eq.~\eqref{eq:wavelength}, which explains our findings.

\textit{SSB in confined N-particle systems.}--To investigate the influence of dielectric nanoconfinement on the collective behavior of quasi-2D charged systems, we further developed a collection of numerical techniques to efficiently evaluate the Green's function Eq.~\eqref{eq:G_point_charge}. A novel Ewald-splitting type strategy is proposed, together with renormalization techniques and fast convergent quadrature schemes. All fine details and numerical validations are provided in the SI~\cite{SI}.
Our study focuses on a prototypical quasi-2D charged system, consisting of a binary mixture of charged particles described by the primitive model.
The system comprises~$N/2$ cations and~$N/2$ anions, each with the same diameter~$\tau_0$ and valence~$\pm 1$, resulting in an overall charge-neutral system. 
The Hamiltonian of the system is defined as follows, where~$i$ represents the~$i$-th particle with charge~$q_i$ located at position~$\V{r_i}$:
\begin{equation}
   \mathcal H = \frac{1}{2} \sum_{i,j=1}^{N}{}^\prime q_i q_j \ell_B G(\V r_i, \V r_j) + U_{\mathrm{LJ}}\;,\label{eq:Hamiltonian}
\end{equation}
The sum notation~$\sum_{i,j}{}^\prime$ implies that when~$i=j$, the function~$G(\V r, \V r)$ corresponds to the self-interaction term, and~$U_{\mathrm{LJ}}$ is the shift-truncated Lennard-Jones (LJ) potential energy used to model excluded-volume interactions. 
While this model disregards other important interactions observed in experimental realizations, it enables us to isolate the dielectric confinement effect. 
Similar systems have been studied recently in Refs.~\cite{dos2017simulations,liang2020harmonic,yuan2021particle}.


\begin{figure}
	\centering
	\includegraphics[width=0.43\textwidth]{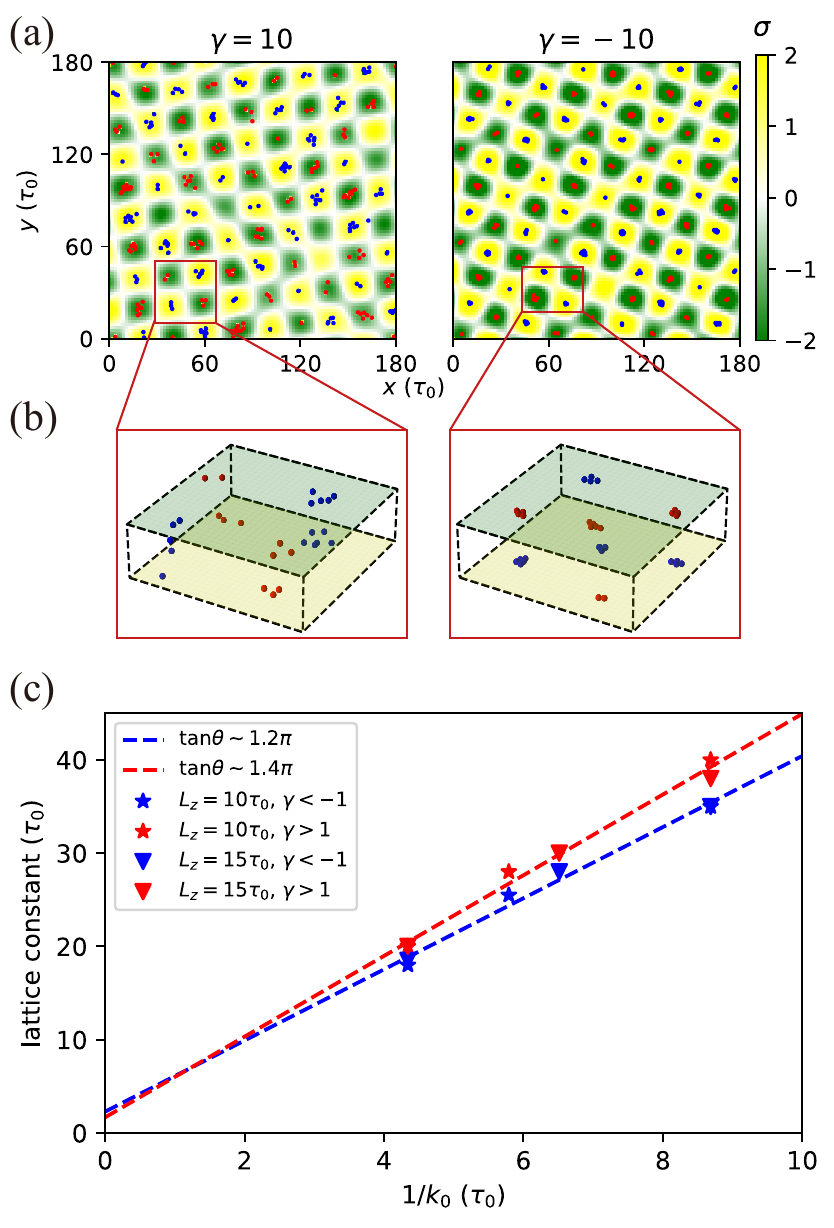}
	\caption{\label{fig:MD} 
        (a): Global particle distributions near the lower substrate and induced surface charge densities for~$\gamma = \pm 10$ and~$L_z = 10$. 
        Positive/negative induced surface charges are in yellow/green, while positive/negative particles are in red/blue, respectively.
        $\sigma$ unit:~$e_0/\tau_0^2$.
        (b): local 3D structures of the charged particles, enlarged from (a), while upper/lower boundaries are in green/yellow, respectively.
        (c): numerical validations for the relationship between the lattice constant and $k_0$. Symbols showing data points from individual simulations, dashed lines depict the linear fitted result.}
\end{figure}

In all the MD simulations, we maintain a constant box size in the~$xy$ plane of~$180\tau_0 \times 180\tau_0$, which is confirmed to eliminate boundary effects.
We vary the values of~$L_z$ and~$\gamma$ to adjust the wave number~$k_0$. 
The system contains~$300$ cations and~$300$ anions.
To isolate electrostatic effect, the reduced temperature $T_r$ is defined as~$T_r =k_{\mathrm B}T/\varepsilon_{\mathrm{Coul}}$, where~$\varepsilon_{\mathrm{Coul}} = e_0^2/(4 \pi \eps (3.5 \tau_0))$ and we set~$\varepsilon_{\mathrm{LJ}} = k_B T$ for both particle-particle and particle-substrate interactions. 
We integrate the temporal evolution using the Velocity-Verlet algorithm and control the temperature using the Anderson thermostat with stochastic collision frequency~$\omega = 0.1$ and reduced temperature~$T_r = 1$.

In the $\abs{\gamma}\leq 1$ regime, extensive simulation works have been done recently~\cite{liang2020harmonic,yuan2021particle} and no SSB phenomenon has been found, i.e., the density distributions of cations $\rho_{+}(\V r)$ and anions $\rho_{-}(\V r)$  always maintain symmetries of the system, given by 1) \emph{cross symmetry} in the confined space: $\rho_{+}(\V r)=\rho_{-}(\V r)$, 2) \emph{longitudinal symmetry}: $\rho_{\pm}(x,y,z)=\rho_{\pm}(x,y,L_z-z)$, and 3) \emph{transverse symmetry}: $\rho_{\pm}(x,y,z)=\rho_{\pm}(x',y',z)$. 
Our simulations give symmetric results for  $\abs{\gamma}\leq 1$, consistent as previous investigations (details are documented in SI~\cite{SI}). 
In the following discussions we will focus on the strongly polarizable cases of $\abs{\gamma}>1$, where SSB phenomena arise.

Fig.~\ref{fig:MD}(a) shows two snapshots for particle distributions near the lower substrate and the corresponding induced surface charge densities, for $\gamma=\pm10$ and $L_z=10$. It clearly shows, for the first time, SSB phenomena in such dielectric confined charged system: both the cross and transverse symmetries are broken when $\gamma=10$; and the remaining longitudinal symmetry is further broken when $\gamma=-10$ (as shown in Fig.~\ref{fig:MD}(b)).

Globally, we observe charged particles spontaneously forming square lattice structures near the substrates for both~$\gamma > 1$ and~$\gamma < -1$ cases, which breaks the transverse symmetry. 
\rev{We attribute this to the long-range single particle oscillatory field in the $xy$-plane, which directs particles self-organizing into a \emph{checkerboard} structure, so as to enhance the overall induced charge landscape, which helps confining particles in local potential wells.}
Locally within each lattice site, two different particle structures are observed: for~$\gamma >1$, interfacial liquid phase is formed, while for~$\gamma < -1$, likely-charged particles self-assemble into 2D clusters, both can be understood by the near field behaviors due to a single confined particle, as was discussed and illustrated in Fig.~\ref{fig:force_x} (a).

Interestingly, in the longitudinal direction, we find that the interfacial liquids/clusters on opposing substrates are strongly correlated, i.e., there is a one-to-one “pairing” between the opposing particle structures, as show in Fig.~\ref{fig:MD}(b).
For $\gamma=10$, the longitudinal pairing is between symmetrically charged particles; while for $\gamma=-10$, the pairing becomes anti-symmetric, which further breaks the longitudinal symmetry. The symmetric/anti-symmetric longitudinal paring is due to the induced charge landscape on opposing substrates, it is clearly that for $\gamma=10$, the checkerboard structures would be matched symmetrically, while for $\gamma=-10$ a negative sign is added to the reflection rates, forming anti-symmetric pairs.

Finally, it is worth noting that the formed square lattices can be well-controlled via the single parameter~$k_0$, consistent with our theoretical prediction.
As shown in Figure \ref{fig:MD}(c), the lattice constant of the system is found to be proportional to $k_0^{-1}$, with various choice of $L_z$ and $\gamma$. 
Two slightly different linear relationships are observed, with fitted ratio~$1.2 \pi$ and~$1.4 \pi$ for~$\gamma < -1$ and~$\gamma > 1$ cases, respectively. 
The distance between neighboring clusters is found to be consistent with the second zero point of the induced surface charge density profile due to a single point charge (see subplots of Fig.~\ref{fig:force_x} (a)). The mechanism allows one to efficiently modulate the collective phase of dielectric confined systems.

\textit{Summary and Conclusions.}--Using a newly developed efficient algorithm that permits simulations of dielectric confined quasi-2D charged systems, we are able to extensively explore the role of dielectric confinement effect.
For a prototypical charge and size symmetric binary particle system, it is discovered for the first time that spontaneous symmetry breaking can be induced and even modulated via the substrate permittivity alone. 
The mechanism of SSB is carefully analyzed, with simple quantitative relation discovered in predicting the formed structures, which provides new physical insights and has potential in future nanodevice design. \rev{While this work discovers the dielectric confinement induced SSB structures, an interesting question remain unanswered is the critical behavior associated with it. According to the Mermin-Wagner theorem~\cite{Mermin1996absence}, it is understood that for 2D systems, continuum symmetry can not be broken spontaneously, for quasi-2D systems studied here, whether it is a first-order or Kosterlitz–Thouless (KT) transition~\cite{Canova2014KT}? This question remains open and to be carefully examined.} 
Our approach also provides a powerful tool for efficient and accurate simulation for a broad range of quasi-2D systems, with wide applications in soft matter physics and advanced materials. 
\rev{Future plans include exploration of the critical behavior of dielectric confined systems, fast algorithm for large-scale simulations~\cite{jin2021random,maxian2021fast}, and its extension to systems with 1) quasi-1D geometry modeling charged nanopores~\cite{Bombardelli2021Electroosmotic}; and 2) tensorial dielectric constants modeling charges under aqueous nanoconfinements~\cite{Loche2020universal}.}

\section*{Supplementary Material}

\rev{
    See the supplementary material for detailed derivation, numerical quadrature scheme, its error analysis, and numerical validations of the method to calculation the electrostatic interaction, as well as the videos of the MD trajectories. 
}

\section*{Acknowledgements}
Z. Gan acknowledges financial support the Natural Science Foundation of China (Grant No. 12201146); Natural Science Foundation of Guangdong (Grant No. 2023A1515012197); Basic and applied basic research project of Guangzhou (Grant No. 2023A04J0054); and Guangzhou-HKUST(GZ) Joint research project (Grant Nos. 2023A03J0003 and 2024A03J0606).
Z. Gan also wish to thank Aleksandar Donev and Leslie Greengard for helpful discussions on the physics and algorithms for quasi-2D charged systems. X. Gao wish to thank Jiuyang Liang, Hongchao Li and Mingzhe Li for fruitful discussions. \rev{Both authors would like to acknowledge the referees for the helpful suggestions and comments, which significantly improves this manuscript.}

\section*{Data Availability}
The data that support the findings of this study are available within the article and its supplementary material.

\bibliography{groupbib}

\end{document}